# Vers une mesure non destructive de la qualité des bois de lutherie

**Marc François**

*LMT Cachan,*
*61 avenue du Président Wilson,*
*94235 Cachan CEDEX*

RESUME. *Une mesure ultrasonore du tenseur d'élasticité d'un épicéa de lutherie dont on fait varier la teneur en eau est présentée. Il est montré qu'un indicateur scalaire connu décrit bien l'évolution de la qualité du bois et que la distance du tenseur d'élasticité au groupe de symétrie tétragonal permet de quantifier la qualité des mesures.*
ABSTRACT. *An ultrasonic measurement of the stiffness tensor of a spruce specimen is made for different moisture content. We show that a known scalar indicator describes correctly the quality of the wood and that the distance from the stiffness tensor to the tetragonal symmetry group describes the quality of the measurements.*
 MOTS-CLES. *mesure ultrasonore - tenseur d'élasticité - épicéa.*

## 1. Introduction

La qualité d'un bois de lutherie est actuellement déterminée empiriquement par les luthiers et les musiciens. Cette étude s'inscrit dans le cadre d'une détermination scientifique de cette qualité. Une appareil, le Lucchi, donne un module d'Young par la mesure ultrasonore d'une célérité entre deux points sur un instrument complet ; c'est une mesure à l'échelle de la structure, intégrant de ce fait la forme et les traitements de surface de l'instrument ou de la planche mesurée.

Les mesures sont ici réalisées à l'échelle du matériau, sur une éprouvette associée à la notion de volume élémentaire représentatif. Elles sont destructives, car il faut usiner une forme spéciale pour les obtenir. Ce sont des mesures ultrasonores qui permettent de déterminer simplement les neuf constantes du tenseur d'élasticité orthotrope [BUC, 88], groupe de symétrie naturellement associé à la microstructure du bois. Les résultats obtenus sont cohérents avec les bases de données existantes à ce sujet [HEA, 48 ; GUI, 88]. Les mesures mécaniques des matériaux anisotropes [BOE, 94a] nécessitent la réalisation d'éprouvettes spécifiques et sont de fait inadaptées à l'objectif à long terme de mesure non destructive sur instruments complets.

Le tenseur d'élasticité est le représentant mathématique le plus global pour décrire l'élasticité linéaire d'un corps anisotrope. Néanmoins, la qualité de la réponse d'un matériau à une sollicitation mécanique est corrélée à certaines pondérations des termes d'élasticité par la masse volumique du matériau [ASH, 80] que nous extrapolerons au tenseur d'élasticité à partir de sa forme la plus simple (en $E/\rho$). Mais ces résultats sont encore difficiles à interpréter, car il faut comparer entre eux des éléments à neuf dimensions.

Un indicateur de qualité doit être scalaire pour permettre le classement entre différents matériaux. Nous vérifierons que la norme du tenseur d'élasticité massique n'est pas un bon indicateur de qualité mais que le critère proposé par différents auteurs [BAR, 97 ; HAI, 79 ; MEY, 95 ; WEG, 96] (équation 4.2) permet effectivement un classement cohérent, au moins pour le matériau étudié.

D'autre part, nous avons exploré la voie de l'analyse des symétries du tenseur d'élasticité [FRA, 98]. Elle permet de définir la distance entre le tenseur d'élasticité mesuré, supposé orthotrope, et les groupes de symétrie plus riche possibles pour ces tenseurs, c'est à dire isotrope transverse, tétragonal, trigonal, cubique et isotrope. Comme l'eau présente au sein du matériau est de symétrie isotrope, elle ne peut pas modifier le groupe de symétrie tétragonal théorique pour ce bois. La distance au tétragonal peut donc être considéré comme un indicateur de la qualité des mesures.

Toutes ces analyses ont étées faites sur un morceau d'épicéa de qualité lutherie fourni par la Cité de la Musique. Sa teneur en eau a été contrôlée à l'aide de l'enceinte climatique Servathin du LMT-Cachan. Nous avons observé que les états de haute teneur en eau, sans doute obtenus de façon trop brutale, ont provoqué un endommagement du bois dont une micrographie est montrée.

## 2. Mesures ultrasonores par contact direct

### 2.1. *Le principe*

Il existe deux principes de mesures ultrasonores : par contact direct et par immersion. Les premières consistent à mesurer la célérité d'une onde ultrasonore à l'aide de deux transducteurs «collés» sur l'échantillon. C'est une manipulation simple et qui a l'intérêt de ne pas privilégier une direction de l'espace. Elle nécessite des éprouvettes de forme globalement cubique et est souvent pratiquée sur le bois [BUC, 92]. Elle est d'autre part la seule à pouvoir donner les 21 constantes du tenseur d'élasticité triclinique (cas le plus général) [FRA, 96].

L'autre méthode est celle par immersion [AUL, 73 ; BAS, 90 ; CAS, 90 ; DIT, 93 ; HAR, 85], elle consiste à mesurer le temps de vol d'une onde ultrasonore émise dans un fluide, réfractée dans l'échantillon qui forme un angle avec la direction de l'émission, et reçue par un second transducteur. Un traitement du problème inverse permet de retrouver le tenseur d'élasticité. Cette dernière méthode fournit généralement des mesures plus précises mais privilégie, par l'utilisation d'un échantillon de forme plaque, certaines directions de l'espace.

### 2.2. *La théorie*

Nous ne rappelons dans cette partie que les équations nécessaires à la compréhension de la suite de ce document. Les démonstrations des principes de base sont détaillées dans les ouvrages génériques [DIE, 74 ; ROS, 99]. La théorie sur les ondes planes est seule nécessaire : on peut mettre en évidence par des mesures laser la cohérence du champ de déplacement de la surface plane d'un transducteur ultrasonore [ROY, 89]. Nous explicitons ci-dessous la méthode de mesure suivant six directions de propagation qui permet d'accéder à l'ensemble des constantes d'élasticité. Elle n'est pas minimale au sens du nombre de mesures (elles sont nombreuses à être redondantes) mais au sens de la simplicité des échantillons.

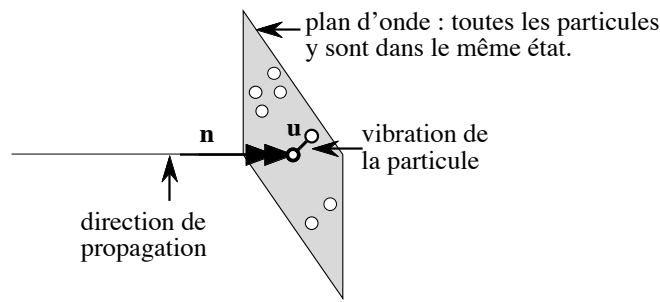

**Figure 2.1.** *Propagation des ondes planes*

Le tenseur acoustique ou tenseur de Christoffel $\Gamma$, du second ordre, est relié au tenseur d'élasticité **C** du matériau et à la direction de propagation **n** de l'onde plane :

$$\Gamma = \mathbf{n}.\mathbf{C}.\mathbf{n} \qquad [2.1]$$

D'autre part, les valeurs propres de $\Gamma$ sont égales aux termes $\rho(V_i)^2$, dans lesquels $\rho$ désigne la masse volumique et $V_i$ les *vitesses de phase* de l'onde plane. La vitesse de phase est celle à laquelle se propage l'information entre les plans d'onde ; elle est donc mesurée suivant **n**. Il convient de la distinguer de la vitesse d'énergie et l'on peut montrer que c'est la vitesse de phase qui intervient dans la mesure ultrasonore par contact direct [ART, 93]. L'indice "i" correspond donc à chacune des trois *polarisations* possibles. Les vecteurs propres de $\Gamma$ correspondent aux directions de vibrations $\mathbf{u}_i$ des particules. Comme $\Gamma$ est symétrique (par construction, à cause des symétries indicielles de **C**), ses vecteurs propres $\mathbf{u}_i$ forment un trièdre orthogonal. Notons au passage que $\Gamma$ est défini positif est donc que les vitesses $V_i$ sont réelles. Au final nous pouvons réécrire $\Gamma$ sous la forme suivante, dans laquelle le symbole $\otimes$ représente le produit tensoriel :

$$\Gamma = \sum_{i=1}^{3} \rho(V_i)^2 \, \mathbf{u}_i \otimes \mathbf{u}_i \qquad [2.2]$$

Ce sont ces deux expressions de $\Gamma$ (équations 2.1 et 2.2) qui permettent d'identifier **C** à partir de la mesure des $\rho(V_i)^2$, la connaissance des **n** et certaines informations sur $\mathbf{u}_i$. L'onde *quasi-longitudinale* est celle qui est la plus proche de la direction de propagation ($\mathbf{u}_i.\mathbf{n}$ est maximal) ; les deux autres sont les directions *quasi-transversales*. Lorsque l'onde quasi-longitudinale est exactement longitudinale on la nomme une onde pure (les polarisations quasi-transverses sont alors elles aussi exactement transverses). On peut démontrer [JAR, 94] qu'une onde pure existe dans le cas ou la direction de propagation est orthogonale à un plan de symétrie. Ce cas se produira, pour notre symétrie orthotrope, pour les propagations suivant les directions $\mathbf{e}_1$, $\mathbf{e}_2$, $\mathbf{e}_3$ de notre échantillon taillé dans ses *axes naturels* (figure 3.1).

Les équations (2.1 et 2.2) et la propriété énoncée ci-dessus nous amènent à considérer en premier ces trois cas où $\mathbf{n} \in (\mathbf{e}_1, \mathbf{e}_2, \mathbf{e}_3)$. Notons $V_{pq}$ la vitesse de l'onde se propageant suivant $\mathbf{e}_p$ de direction de vibration $\mathbf{e}_q$. $V_{pp}$ est par conséquent une onde longitudinale et $V_{pq}$ transversale si q≠p. Les tenseurs acoustiques sont diagonaux dans ce cas ; il vient (en notant $C_{IJ}$ l'écriture de Voigt du tenseur $C_{ijkl}$) :

$\mathbf{n} = \mathbf{e}_1$ ; $C_{11} = \rho(V_{11})^2$
$\mathbf{n} = \mathbf{e}_2$ ; $C_{22} = \rho(V_{22})^2$
$\mathbf{n} = \mathbf{e}_3$ ; $C_{33} = \rho(V_{33})^2$
$\mathbf{n} = \mathbf{e}_2, \mathbf{e}_3$ ; $C_{44} = \rho(V_{23})^2 = \rho(V_{32})^2$
$\mathbf{n} = \mathbf{e}_1, \mathbf{e}_3$ ; $C_{55} = \rho(V_{13})^2 = \rho(V_{31})^2$
$\mathbf{n} = \mathbf{e}_1, \mathbf{e}_2$ ; $C_{66} = \rho(V_{12})^2 = \rho(V_{21})^2$ [2.3]

Nous avons dès à présent six constantes matériaux identifiées dont trois ($C_{44}$, $C_{55}$, $C_{66}$) avec une redondance utile pour discerner le «bon» signal transverse parmi les sources d'erreur toujours possibles. Il nous reste à déterminer les termes hors diagonale ($C_{12}$, $C_{23}$, $C_{31}$). Considérons les mesures suivant les directions de propagations bissectrices des directions naturelles ([$\mathbf{e}_2$,$\mathbf{e}_3$], [$\mathbf{e}_3$,$\mathbf{e}_1$], [$\mathbf{e}_1$,$\mathbf{e}_2$]). Nous les nommons, en s'inspirant de la notation de Voigt, ($e_4,e_5,e_6$). Considérons, par exemple $e_6$ : il n'est pas vecteur propre de $e_6.C.e_6$, tandis que $e_3$ l'est, indiquant une onde transverse pure (elle est orthogonale au plan de symétrie [$e_3^\perp$]). Les deux autres ondes ne sont pas pures (directions de vibration quasi-longitudinale $u_{6l}$ et quasi-transverse $u_{6t}$ dans la figure 2.2) et la mesure ne permet pas la connaissance des directions de vibrations des particules $u_i$.

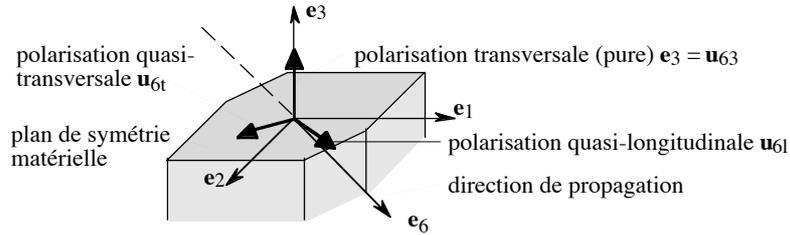

**Figure 2.2.** *Cas de la propagation suivant $e_6$*

On ne peut donc plus égaler directement les deux expressions des tenseurs acoustiques. On considère alors le sous espace (le plan) [$e_3^\perp$]. Les deux tenseurs acoustiques sont égaux en projection sur [$e_3^\perp$], on égalise les déterminants des projections de ces deux tenseurs. Il vient, en nommant $V_{6l}$ la vitesse quasi-longitudinale et $V_{6t}$ la vitesse quasi transversale dans le plan [$e_3^\perp$] :

$$\mathbf{n} = \mathbf{e}_4 \,;\, C_{23} = \sqrt{(C_{22}+C_{44})(C_{33}+C_{44}) - 4\rho^2 (V_{4l})^2 (V_{4t})^2} - C_{44}$$
$$\mathbf{n} = \mathbf{e}_5 \,;\, C_{31} = \sqrt{(C_{33}+C_{55})(C_{11}+C_{55}) - 4\rho^2 (V_{5l})^2 (V_{5t})^2} - C_{55}$$
$$\mathbf{n} = \mathbf{e}_6 \,;\, C_{12} = \sqrt{(C_{11}+C_{66})(C_{22}+C_{66}) - 4\rho^2 (V_{6l})^2 (V_{6t})^2} - C_{66} \quad [2.4]$$

L'autre équation déduite de (2.1 et 2.2) pour $\mathbf{n}=\mathbf{e}_6$ en projection sur le sous-espace $\mathbf{e}_3$ permet de déterminer une seconde fois la somme $C_{44}+C_{55}$. La mesure de cette vitesse transverse sera utilisable afin de vérification de mesures précédentes. Nous pouvons résumer ces mesures redondantes sur les trois bissectrices par :

$$\mathbf{n} = \mathbf{e}_4 \,;\, \frac{C_{55}+C_{66}}{2} = \rho\, (V_{41})^2$$
$$\mathbf{n} = \mathbf{e}_5 \,;\, \frac{C_{66}+C_{44}}{2} = \rho\, (V_{52})^2$$
$$\mathbf{n} = \mathbf{e}_6 \,;\, \frac{C_{44}+C_{55}}{2} = \rho\, (V_{63})^2 \quad [2.5]$$

Au bilan, nous pouvons détailler la campagne de mesures et justifier la forme des éprouvettes :

— sur une éprouvette cubique, les trois directions de propagation $e_1$, $e_2$, $e_3$ nous fournissent les modules diagonaux par les équations (2.3). Les termes du second quadrant $C_{44}$, $C_{55}$, $C_{66}$ sont obtenus deux fois.

— sur trois éprouvettes possédant une paire de faces orthogonales aux bissectrices $e_4$, $e_5$, $e_6$ nous mesurons les termes non diagonaux $C_{23}$ $C_{31}$, $C_{12}$ par les équations (2.4) et une nouvelle mesure des termes du second quadrant par les équations (2.5).

Nous avons bien sûr utilisé ces redondances pour identifier sur les signaux expérimentaux les diverses polarisations qui sont souvent mélangées les unes avec les autres (chapitre 3). Toute la difficulté expérimentale se trouve dans l'identification des différents signaux.

### 2.3. *Calculs d'erreur*

Nous identifions par le terme $\Delta x$ la marge d'erreur sur la mesure $x \pm \Delta x/2$. Nous regroupons dans ce calcul les différentes mesures de vitesse, censées avoir la même précision $\Delta V$, bien que dans la pratique, certains signaux aient une marge de confiance plus élevée que d'autres. La masse volumique $\rho$ est classiquement obtenue par un rapport masse sur volume d'où :

$$\frac{\Delta \rho}{\rho} = \frac{\Delta m}{m} + 3 \frac{\Delta L}{L} \qquad [2.6]$$

Dans notre application sur l'épicéa, les mesures au pied à coulisse ont une précision de 1/50 mm pour une longueur de l'ordre de 20 mm et la masse est mesurée à $10^{-2}$ g pour une masse d'environ 10 g. Soit $\Delta\rho/\rho$ de l'ordre de $4\,10^{-3}$ dont $1\,10^{-3}$ provient de l'erreur en masse et $3\,10^{-3}$ de l'erreur sur la longueur. Au niveau des composantes appartenant à la diagonale du tenseur d'élasticité, les équations (2.3) nous donnent le calcul d'erreur suivant :

$$\frac{\Delta C_{ii}}{C_{ii}} = \frac{\Delta \rho}{\rho} + 2 \frac{\Delta V}{V} \qquad [2.7]$$

En retenant une erreur moyenne en vitesse de 2,5%, nous obtenons une erreur $\Delta C_{ii}/C_{ii}$ de 5%, et l'on constate que l'erreur en masse volumique est tout à fait négligeable à ce niveau. Il serait plus judicieux de considérer à chaque mesure l'erreur commise, qui peut être plus faible pour un signal bien isolé des autres et de bonne intensité. En général, pour les grandes célérités correspondant aux modules $C_{ii}$ les plus forts l'erreur en vitesse peut raisonnablement être estimée à 1% donc 2% sur les modules $C_{ii}$ les plus grands par exemple sur le $C_{11}$ mesuré sur l'épicéa à 16 MPa environ (table 1). A l'opposé, certains modules plus faibles présenteront des signaux beaucoup plus difficiles à distinguer des autres et des phénomènes parasites, et la précision pourra chuter. Une estimation pourrait être de 10% d'erreur en vitesse et donc de 20% sur les modules $C_{ii}$ les plus faibles, comme le $C_{44}$ mesuré pour l'épicéa de l'ordre de 0,1 MPa. Au niveau des composantes hors diagonales, les équations (2.4) nous donnent, par un calcul d'erreur grossier qui suppose les valeurs de $C_{ii}$, $C_{ij}$ et $\rho V^2$ peu différentes :

$$\frac{\Delta C_{ij}}{C_{ij}} = 5 \frac{\Delta \rho}{\rho} + 10 \frac{\Delta V}{V} \qquad [2.8]$$

En retenant l'estimation globale de l'erreur moyenne en vitesse de 2,5% soit $\Delta C_{ii}/C_{ii}$ de 5%, nous obtenons pour ces composantes $C_{ij}$ une erreur $\Delta C_{ij}/C_{ij}$ de l'ordre de 25%, une précision de l'ordre de cinq fois moins bonne que celle des $C_{ii}$. Bien que cette valeur élevée soit un majorant, il apparaît que les termes hors diagonale sont obtenus avec moins de précision que ceux de la diagonale.

## 3. Dispositif expérimental

### 3.1. *Les éprouvettes*

L'étude présentée dans les chapitres précédents nous impose de réaliser un système d'éprouvettes permettant six directions de propagations ($e_1$, $e_2$, $e_3$, $e_4$, $e_5$, $e_6$). Chaque direction de propagation requiert une paire de faces parallèles. Une solution est de tailler une éprouvette unique dans un cube [NEI, 67] mais elle impose un grand volume de matière du fait que chaque face doit pouvoir contenir l'intégralité de la surface active du transducteur. La dimension "standard" d'une pastille de transducteur étant de diamètre 12,7 mm, il est nous était impossible d'obtenir un cube d'environ 40 mm d'arête dans notre échantillon d'épicéa. La découpe a été faite de façon à respecter au mieux les directions de symétries matérielles du bois ; la direction 1 correspond à la direction longitudinale de l'arbre (1=L), la 2 à la direction radiale (2=R) et la 3 à la direction tangentielle (3=T). Nous avons réalisé quatre éprouvettes : un cube permettant de mesurer les propagations suivant les trois directions principales, les trois autres permettent les propagations suivant les bissectrices représentées en gras dans la figure (3.1). L'usinage a été réalisé avec une machine-outil (fraiseuse) permettant une précision d'$1/100^e$ de mm environ ; toutefois le bois, par sa microstructure, ne permet qu'une précision réaliste d'environ $1/10^e$ de mm suffisante aux vues du calcul d'erreur (chapitre 2.3). La dimension des éprouvettes est conditionnée par la longueur d'onde. Nos transducteurs sont à 1 MHz, mais le bois se comporte en filtre et le signal reçu possède une fréquence de l'ordre de 0.25 MHz. La célérité la plus faible mesurée était de l'ordre de $10^3$ m/s ce qui conduit à une longueur d'onde $\lambda$ d'environ 4 mm. Les hétérogénéité correspondent pour le bois à la distance entre les rayons, soit environ 1 mm. Nos éprouvettes ont des dimensions caractéristiques de l'ordre de 20 mm, soit donc environ cinq longueurs d'onde, c'est un compromis entre bonne réception et nombre d'hétérogénéité. Les mesures par contact direct nécessitent un couplant assurant une bonne transmission des ondes ; nous avons utilisé de la graisse silicone. Afin d'éviter la pénétration de celle-ci dans l'épicéa, nous avons recouvert les éprouvettes de ruban adhésif car le vernissage n'aurait pas permis la variation de teneur en eau [CHO, 99].

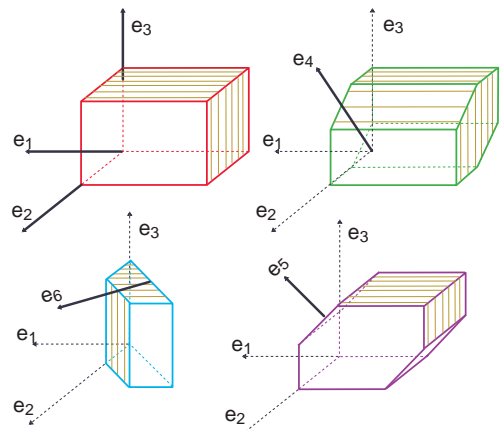

**Figure 3.1.** *Forme des 4 éprouvettes ultrasonores (épaisseur env. 25mm)*

La mesure de la masse volumique est faite par le rapport masse/volume. La masse est mesurée par une balance de précision au $1/100^e$ g et le volume par mesure de longueur au pied à coulisse. La variation de la teneur en eau a été réalisée au sein d'une enceinte climatique Servathin sauf pour la saturation obtenue par immersion pendant plusieurs jours. Une source d'imprécision demeure pourtant du fait de la plus rapide diffusion de l'eau pour les géométries en coin que pour les géométries compactes, ce qui fait que, pour une masse volumique globale donnée, il est possible d'avoir un gradient au sein de l'échantillon. Une mesure plus fine de la teneur en eau serait à envisager pour la suite.

### 3.2. *Le montage*

Nous utilisons une carte "Saphir" [LEC, 93] permettant l'émission d'un "pseudo-dirac" de largeur 100ns et de tension 230V. Cette carte est implantée dans un micro-ordinateur. Elle est munie d'un oscilloscope intégré auquel nous préférons substituer un oscilloscope numérique qui est synchronisé sur l'impulsion fournie dans le câble émetteur. Le montage présenté figure (3.2) est prévu pour les mesures sur éprouvettes à 26 faces relatives aux mesures des matériaux tricliniques ; une masse cylindrique permet de garantir un effort presseur constant. Les transducteurs transversaux dont maintenus parallèles par une paire de bras couplés visibles sur la photographie.

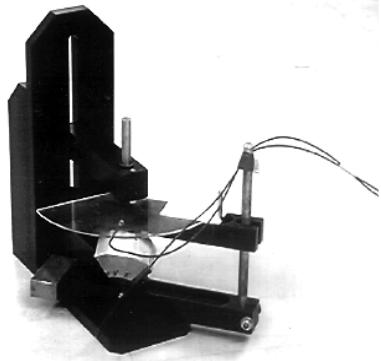

**Figure 3.2.** *Le montage ultrasonore*

### 3.3. *Les mesures*

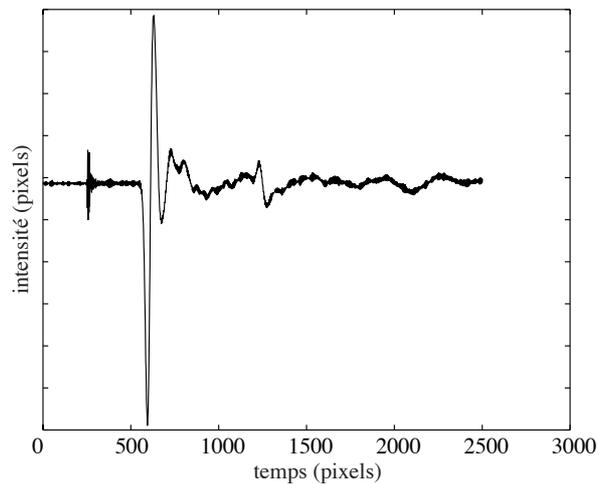

**Figure 3.3.** *Mesure de signaux ultrasonores quasi-longitudinaux*

En regardant la figure (3.3), il apparaît un problème prioritaire : comment définir le point de mesure du temps de vol ? La première impulsion correspond au parasitage de la chaîne par l'impulsion initiale. L'émetteur envoie environ une alternance et demi (charte constructeur) qui suit logiquement l'impulsion. Il semble alors naturel de considérer l'instant à mesurer comme celui du premier événement à la réception. Une mesure sur un cale à deux épaisseurs l'une double de l'autre permet d'ailleurs de vérifier que seul ce point respecte cette homothétie. Néanmoins, la mesure des signaux transverses nécessite de la part de l'opérateur de séparer le signal à mesurer des signaux plus rapides, en général la résiduelle longitudinale.

## 4. Résultats obtenus

### 4.1. *Les tenseurs d'élasticité*

Nous présentons ci-dessous les tenseurs d'élasticité obtenus pour chaque teneur en eau testée, par la méthode décrite au paragraphe 2. Le tenseur est exprimé en contraction de Voigt, c'est à dire que le terme $C_{14}$ est égal au terme $C_{1123}$, par exemple. L'unité est le MPa. La teneur en eau $\tau_{eau}$ est la exprimée comme le rapport masse d'eau / masse de bois sec (soit encore $\rho/\rho_{sec} - 1$).

$$\begin{vmatrix} 16.2 & 1.5 & 1.3 & 0 & 0 & 0 \\ 1.5 & 2.7 & 0.6 & 0 & 0 & 0 \\ 1.3 & 0.6 & 1.7 & 0 & 0 & 0 \\ 0 & 0 & 0 & 0.1 & 0 & 0 \\ 0 & 0 & 0 & 0 & 0.9 & 0 \\ 0 & 0 & 0 & 0 & 0 & 1.0 \end{vmatrix} \qquad \begin{vmatrix} 15.4 & 1.2 & 1.1 & 0 & 0 & 0 \\ 1.2 & 2.7 & 0.7 & 0 & 0 & 0 \\ 1.1 & 0.7 & 1.6 & 0 & 0 & 0 \\ 0 & 0 & 0 & 0.1 & 0 & 0 \\ 0 & 0 & 0 & 0 & 0.9 & 0 \\ 0 & 0 & 0 & 0 & 0 & 1.0 \end{vmatrix}$$

**Tableau 4.1.** *C pour $\tau_{eau} = 0\ \%$*  **Tableau 4.2.** *C pour $\tau_{eau} = 2{,}8\ \%$*

$$\begin{vmatrix} 14.6 & 0.8 & 0.9 & 0 & 0 & 0 \\ 0.8 & 2.5 & 0.4 & 0 & 0 & 0 \\ 0.9 & 0.4 & 1.5 & 0 & 0 & 0 \\ 0 & 0 & 0 & 0.09 & 0 & 0 \\ 0 & 0 & 0 & 0 & 0.7 & 0 \\ 0 & 0 & 0 & 0 & 0 & 0.9 \end{vmatrix} \qquad \begin{vmatrix} 14.3 & 0.6 & 0.8 & 0 & 0 & 0 \\ 0.6 & 2.2 & 0.4 & 0 & 0 & 0 \\ 0.8 & 0.4 & 1.4 & 0 & 0 & 0 \\ 0 & 0 & 0 & 0.09 & 0 & 0 \\ 0 & 0 & 0 & 0 & 0.7 & 0 \\ 0 & 0 & 0 & 0 & 0 & 0.8 \end{vmatrix}$$

**Tableau 4.3.** *C pour $\tau_{eau} = 4{,}6\ \%$*  **Tableau 4.4.** *C pour $\tau_{eau} = 7{,}2\ \%$*

$$\begin{vmatrix} 13.8 & 0.3 & 1.2 & 0 & 0 & 0 \\ 0.3 & 1.4 & 0.7 & 0 & 0 & 0 \\ 1.2 & 0.7 & 1.0 & 0 & 0 & 0 \\ 0 & 0 & 0 & 0{,}04 & 0 & 0 \\ 0 & 0 & 0 & 0 & 0.5 & 0 \\ 0 & 0 & 0 & 0 & 0 & 0.7 \end{vmatrix} \qquad \begin{vmatrix} 14.6 & 1.9 & 1.5 & 0 & 0 & 0 \\ 1.9 & 2.1 & 1.2 & 0 & 0 & 0 \\ 1.5 & 1.2 & 1.2 & 0 & 0 & 0 \\ 0 & 0 & 0 & 0{,}04 & 0 & 0 \\ 0 & 0 & 0 & 0 & 0.5 & 0 \\ 0 & 0 & 0 & 0 & 0 & 0.7 \end{vmatrix}$$

**Tableau 4.5.** *C pour $\tau_{eau}$ = 11,6 %*      **Tableau 4.6.** *C pour $\tau_{eau}$ = 16,0 %*

$$\begin{vmatrix} 15.9 & 2.6 & 1.8 & 0 & 0 & 0 \\ 2.6 & 2.2 & 1.3 & 0 & 0 & 0 \\ 1.8 & 1.3 & 1.3 & 0 & 0 & 0 \\ 0 & 0 & 0 & 0{,}04 & 0 & 0 \\ 0 & 0 & 0 & 0 & 0.6 & 0 \\ 0 & 0 & 0 & 0 & 0 & 0.8 \end{vmatrix} \qquad \begin{vmatrix} 16.4 & 2.9 & 2.1 & 0 & 0 & 0 \\ 2.9 & 2.0 & 1.2 & 0 & 0 & 0 \\ 2.1 & 1.2 & 1.1 & 0 & 0 & 0 \\ 0 & 0 & 0 & 0{,}03 & 0 & 0 \\ 0 & 0 & 0 & 0 & 0.6 & 0 \\ 0 & 0 & 0 & 0 & 0 & 0.7 \end{vmatrix}$$

**Tableau 4.7.** *C pour $\tau_{eau}$ = 24,7 %*      **Tableau 4.8.** *C pour $\tau_{eau}$ = 31,3 %*

$$\begin{vmatrix} 16.6 & 2.8 & 2.1 & 0 & 0 & 0 \\ 2.8 & 2.1 & 1.2 & 0 & 0 & 0 \\ 2.1 & 1.2 & 1.2 & 0 & 0 & 0 \\ 0 & 0 & 0 & 0{,}03 & 0 & 0 \\ 0 & 0 & 0 & 0 & 0.6 & 0 \\ 0 & 0 & 0 & 0 & 0 & 0.7 \end{vmatrix}$$

**Tableau 4.9.** *C pour $\tau_{eau}$ = 33,9 %*

Ces résultats sont présentés sous forme de graphe (figure 4.1) dans lequel la chronologie de la mesure est présentée par un numéro d'état. Ils peuvent déconcerter au premier abord au niveau des grandes valeurs obtenues pour les fortes teneur en eau. Il est cependant plus naturel de considérer la rigidité spécifique du matériau, pour laquelle le tenseur d'élasticité est divisé par la masse volumique (figure 4.2).

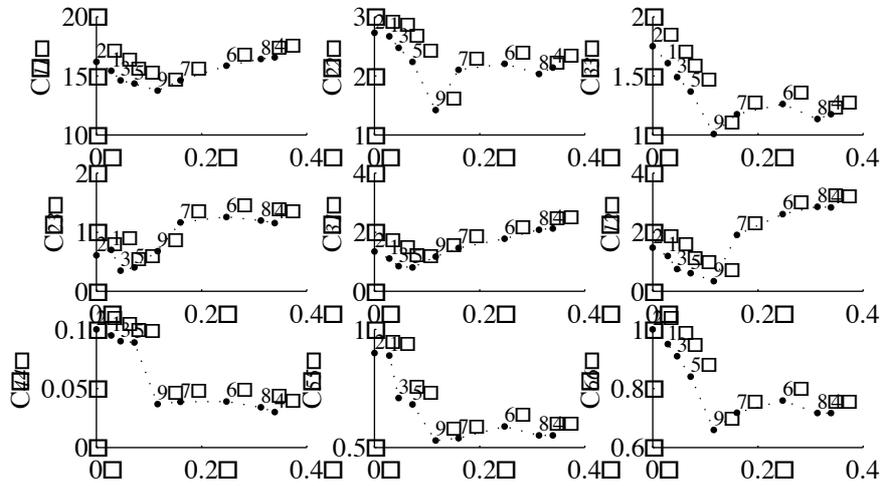

**Figure 4.1.** *Tenseur d'élasticité en fonction de la teneur en eau*

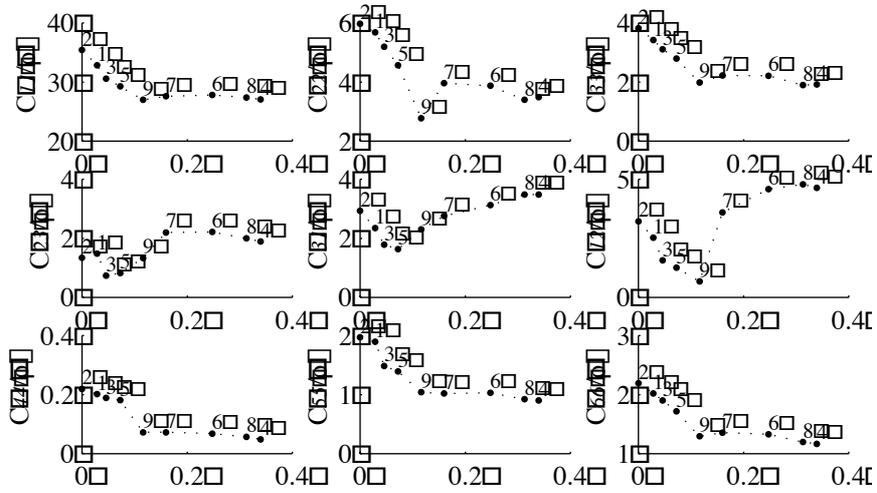

**Figure 4.2.** *Tenseur massique en fonction de la teneur en eau*

Nous pouvons observer la décroissance monotone des termes $C_{ii}/\rho$. Ceci entraîne une conclusion immédiate qui sera vérifiée ensuite : *l'état optimal pour le bois de lutherie est l'état sec*. L'évolution des termes $C_{23}/\rho$, $C_{31}/\rho$ et $C_{12}/\rho$ présente pourtant la caractéristique de décroître puis croître à nouveau pour les teneurs en eau supérieures à 10% environ. Ce phénomène s'explique classiquement en mécanique des sols : un terme comme $C_{23}$ représente la réponse en contrainte $\sigma_{22}$ déformation imposée $\varepsilon_{33}$ ; or ce mode de déformation est non isochore (entraîne un changement de volume) et l'eau, qui possède un haut module de compressibilité, y participe activement. Néanmoins, les termes «importants» en vibration de plaques ou coques sont essentiellement les $C_{11}/\rho$ et $C_{22}/\rho$ (directions L et R de découpe des planches), ce qui valide donc la conclusion précédente. La singularité observée pour l'état(9) sera expliquée au chapitre (4.2) par l'existence d'un endommagement.

### 4.2. Observation d'un endommagement

Lors de l'expérimentation, les traitements sous enceinte climatique ont fait apparaître, dès l'état (5) et de façon de plus en plus marquée jusqu'à l'état (9) un réseau de fissures visibles sur la figure (4.3). Ces fissures, perpendiculaires aux rayons du bois, sont de taille millimétrique. Leur création est sans doute liée à une pression de vapeur générée lors du séchage sans doute trop rapide de l'échantillon. Néanmoins, cette observation peut guider vers différends travaux de modélisation à l'aide de la théorie de l'endommagement anisotrope [LEM, 92]. D'autre part, il faut considérer la confiance des mesures postérieures à l'état (5) comme moins bonne. L'état (9) se distingue d'ailleurs visiblement des autres sur l'ensemble des courbes présentées par des valeurs plus faibles liées à cet endommagement.

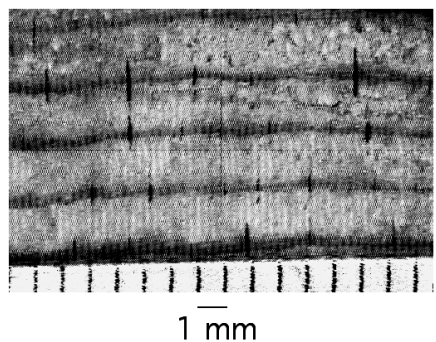

**Figure 4.3.** *Micrographie de l'épicéa endommagé*

### 4.3. Indicateurs scalaires de la qualité du matériau

Les résultats ci-dessus ont encore l'inconvénient d'être à neuf dimensions. Une comparaison entre différents matériaux, notre objectif à long terme n'est possible que sur un indicateur scalaire. Une voie en cours d'exploration [HOU, 99] est celle d'une pondération entre les différends $C_{ij}/\rho$ permettant de retrouver les classements effectués subjectivement par différend luthiers. Une autre est de rechercher un indicateur plus «mathématique». La notion la plus simple est celle de la norme euclidienne naturelle qui vaut, pour un tenseur d'élasticité spécifique $\mathbf{C}/\rho$ :

$$\| \mathbf{C}/\rho \| = \frac{1}{\rho}\sqrt{C_{ijkl}\,C_{ijkl}} \qquad [4.1]$$

Le graphe obtenu, représenté figure (4.4) montre que ce critère n'est pas sélectif pour les hautes teneurs en eau et, peut être trop pour les basses (les musiciens ne recherchent pas systématiquement une ambiance parfaitement sèche pour leurs instruments).

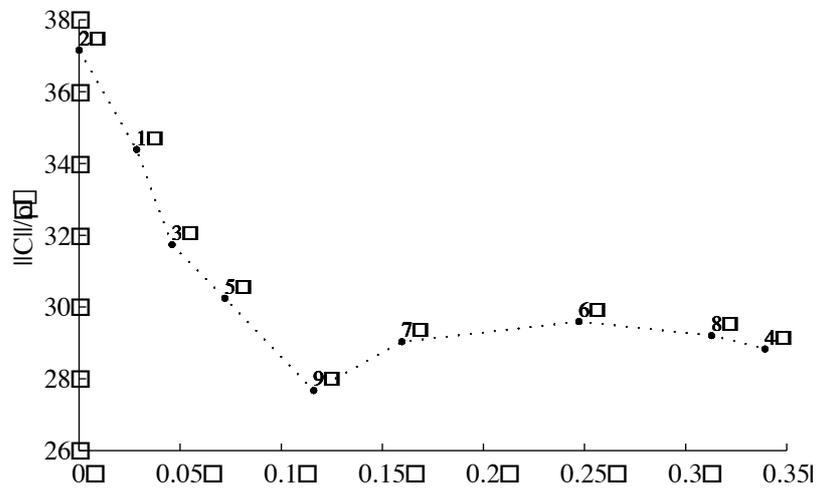

**Figure 4.4.** *Norme de $\mathbf{C}/\rho$ en fonction de la teneur en eau*

Différents auteurs [BAR, 97 ; HAI, 79 ; MEY, 95 ; WEG, 96], montrent l'importance du terme statique suivant (équation 4.2) pour la qualité de la résonance d'une plaque d'instrument de musique. Les termes $E_L$ et $E_R$ désignent respectivement les modules d'Young dans les directions longitudinales et radiales correspondant à la découpe des planches du bois dans notre cas.

$$M = \frac{(E_L\,E_R)^{(1/4)}}{\rho^{(3/2)}} \qquad [4.2]$$

Nous rappelons l'expression du module d'Young dans une direction $\mathbf{n}$ donnée d'un tenseur d'élasticité $\mathbf{C}$ quelconque :

$$E(\mathbf{n}) = \frac{1}{(\mathbf{n} \otimes \mathbf{n}) : \mathbf{C}^{-1} : (\mathbf{n} \otimes \mathbf{n})} \qquad [4.3]$$

La figure (4.5) suivante montre les différentes valeurs obtenues pour le terme M pour chaque teneur en eau, à partir des tables (4.1 à 4.9) ci-dessus.

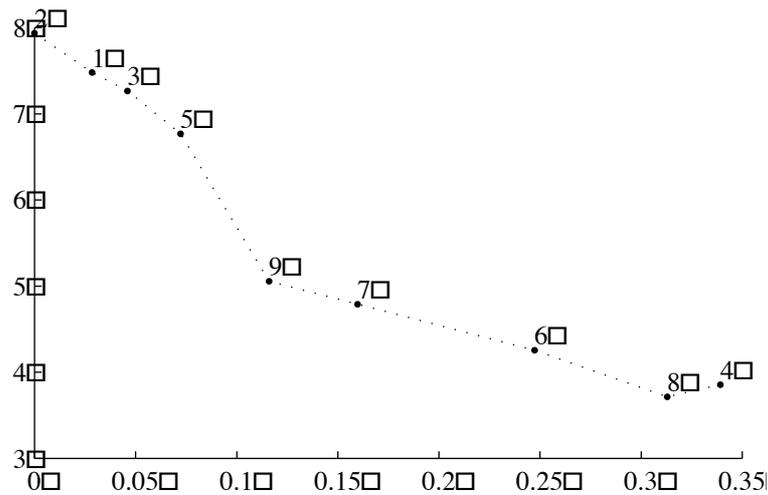

**Figure 4.5.** *Critère M en fonction de la teneur en eau*

La décroissance quasi-monotone de la qualité en fonction de la teneur en eau est bien visible. Les états (1) à (3) à basse teneur en eau sont par contre peu distincts, au contraire de la représentation de la norme du tenseur d'élasticité **C** pondérée (figure 4.4), ce qui est cohérent avec l'expérience. La sélection par M est aussi plus «contrastée» au sens de l'écart entre les plus hautes et basses valeurs (de 4 à 8) par rapport à la mesure en norme de **C**/ρ (de 28 à 38). Ces considérations amènent à penser qu'il s'agit d'un «bon» représentant scalaire de la qualité du matériau.

### 4.3. *Indicateur scalaire de la qualité des mesures*

L'autre voie explorée est celle de la distance entre le tenseur d'élasticité et les groupes de symétrie plus riches, au sens de l'arborescence des symétries des tenseurs de ce type [YON, 91]. Les détails de la théorie permettant de déterminer le tenseur possédant une symétrie donnée le plus proche d'un autre peuvent être trouvés dans la littérature [FRA, 98]. La distance entre un tenseur d'élasticité de référence **C** et un autre **D** est définie à partir de la norme euclidienne naturelle (équation 4.1) comme :

$$d(\mathbf{C},\mathbf{D}) = \frac{\|\mathbf{C} - \mathbf{D}\|}{\|\mathbf{C}\|} \qquad [4.2]$$

Le dessin au sein de chaque case de la figure (4.6) représente la trace des plans de symétrie sur un cube pour le niveau de symétrie considéré, pour mémoire. Il apparaît que le tenseur mesuré est très proche (4.2 %) d'être tétragonal, et donc de pouvoir être décrit par 6 constantes d'élasticité au lieu de neuf. Ce groupe de symétrie, associé à une maille parallélépipèdique à base carrée, est d'ailleurs compréhensible au vu de la microstructure de ce type de bois [GAU, 80].

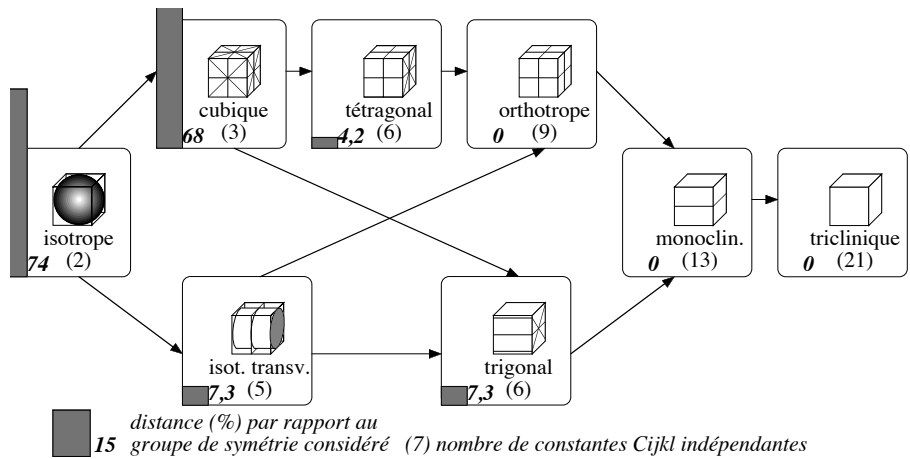

**Figure 4.6.** *Distance aux différents groupes de symétrie pour l'épicéa sec*

Compte tenu de ces remarques, il est naturel de penser que la distance au tétragonal puisse être aussi un indicateur de qualité. Le graphe (figure 4.7) montre qu'il n'y a pas de corrélation avec la teneur en eau, ni avec la qualité du matériau indiquée par la figure (4.5). Considérant que la symétrie tétragonale est naturelle pour ce matériau et que l'eau, de symétrie isotrope, ne peut la modifier par sa présence, on en déduit que cette distance au tétragonal est un bon représentant de la *qualité des mesures*. La mesure de l'état sec (2), de meilleure qualité, est d'ailleurs la plus simple pour l'expérimentateur du fait de l'absence des difficultés liées à la constance de l'hygrométrie au sein de l'éprouvette. La mesure à l'état initial (1) présente elle aussi un niveau de qualité comparable. Par contre les états de haute teneur en eau ou finaux (matériau endommagé) présentent une moins bonne qualité de mesure. La distance au tétragonal est dans notre cas proportionnelle aux différences $|C_{22} - C_{33}|$, $|C_{55} - C_{66}|$, $|C_{12} - C_{13}|$ ; cette dernière est une condition sur les termes hors diagonale obtenus avec moins de précision (chapitre 2.3).

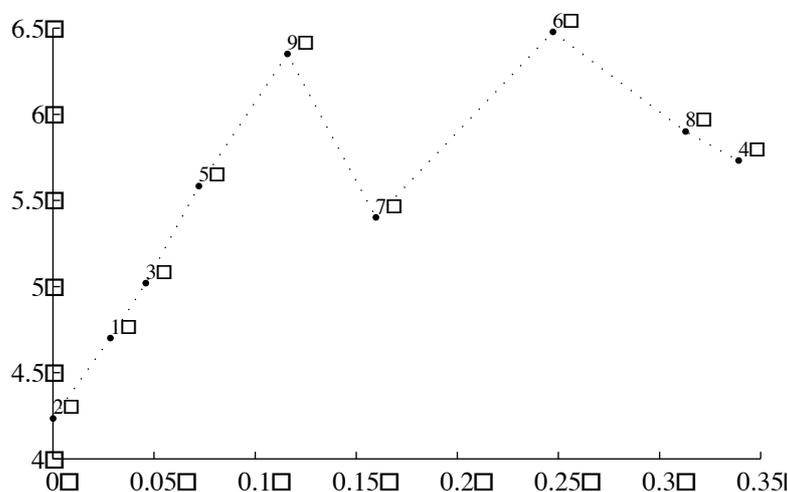

**Figure 4.7.** *Distance au tétragonal en fonction de la teneur en eau*

Enfin, d'autres indicateurs pourraient être tirés de l'analyse tensorielle, comme les décompositions de Boehler ou de Kelvin [BOE, 94b ; RYC, 84].

**5. Conclusions et perspectives**

Les mesures ultrasonores sont un procédé qui peut être non destructif, même si l'identification de toutes les constantes d'élasticité par méthode non destructive n'est pas encore possible à ce jour. Les mesures par immersion semblent plus adaptées à cette approche que les mesures par contact direct réalisées dans cette étude mais l'appareillage et le traitement des données restent à établir pour une structure telle qu'un instrument de musique.

La comparaison en terme de qualité de deux matériaux distincts (bois, composites, métaux…) semble possible par l'utilisation d'un indicateur scalaire même si les travaux présentés demandent à être confirmés sur une base de donnée plus étendue et comparé à l'évaluation intuitive de spécialistes. La distance du tenseur d'élasticité au groupe de symétrie théorique pour le matériau est un indicateur de la qualité des mesures, au sens ou il vérifie (en partie) la cohérence des constantes trouvées.

Ces évaluateurs, qui demandent encore à être validés, peuvent être étendus à d'autres matériaux de lutherie (métaux, composites…) mais aussi à d'autres domaines utilisant ces matériaux (aviation, génie civil…).

# Bibliographie